\documentclass[prl,aps,twocolumn,superscriptaddress,floatfix]{revtex4}
\usepackage{graphicx}% Include figure files
\usepackage{times}
\usepackage{color}
\definecolor{darkblue}{rgb}{0,0.02,0.45}
\usepackage[colorlinks=true,citecolor=darkblue]{hyperref}

\def \mst {{\rm Mn$_3$Si$_2$Te$_6$}}
\def \TC {T_c}

\begin{document}

\title{Magnetic Structure and Spin Fluctuations in Colossal Magnetoresistance Ferrimagnet $\rm Mn_3Si_2Te_6$}

\author{Feng Ye}
\email{yef1@ornl.gov}
\author{Masaaki Matsuda}
\author{Zachary~Morgan}
\author{Todd~Sherline}
\affiliation{Neutron Scattering Division, Oak Ridge National Laboratory,
Oak Ridge, Tennessee 37831, USA}
\author{Yifei Ni}
\author{Hengdi Zhao}
\author{G. Cao}
\affiliation{
Department of Physics, University of Colorado at Boulder, Boulder,
Colorado 80309, USA}

\date{\today}

%\pacs{75.30.Ds, 61.12.Ld, 71.15.Mb}

\begin{abstract}
    The ferrimagnetic insulator \mst, which features a Curie temperature $T_c$ at
    78~K and a delicate yet consequential magnetic frustration,
    exhibits colossal magnetoresistance (CMR) when the
    magnetic field is applied along the magnetic hard axis, surprisingly
    inconsistent with existing precedents [Y. Ni, H. Zhao, Y. Zhang {\it et
    al.} Phys. Rev. B 103, L161105 (2021)]. This discovery
    motivates a thorough single-crystal neutron diffraction study in order
    to gain insights into the magnetic structure and its hidden correlation
    with the new type of CMR.  Here we report a noncollinear magnetic
    structure below the $T_c$ where the moments lie predominantly
    within the basal plane but tilt toward the $c$ axis by $\sim10^{\circ}$ at
    ambient conditions. A substantial magnetic diffuse scattering decays slowly and
    persists well above the $T_c$. The evolution of the spin correlation
    lengths agrees well with the electrical resistivity, underscoring the role
    of spin fluctuation contributing to the magnetoresistivity near the
    transition. Application of magnetic field along the $c$ axis, renders a
    swift occurrence of CMR but only a slow tilting of the magnetic moments
    toward the $c$ axis. The unparalleled changes indicate a non-consequential
    role of magnetic polarization.
\end{abstract}

\maketitle
Colossal magnetoresistance (CMR), the dramatic change in electric resistance in
response to an applied magnetic field, has been known for decades and is
extensively studied in the archetypical perovskite manganites and their 
variants \cite{jin94}. In these systems, the simultaneous insulator-metal
transition and the ferromagnetic order of the mixed Mn$^{3+}$-Mn$^{4+}$
network can be understood in the context of the ``double exchange'' mechanism;
the electron hopping is enhanced when neighboring Mn ions have mutually
aligned spins \cite{zener51,anderson55,degennes60}.
The spin order promotes the electron hopping and increases the
effective exchange interaction, annealing out the lattice distortion, thus
leading to the insulator-metal transition.

Exploring new classes of materials exhibiting CMR beyond the perovskite
manganites continues to inspire huge interest, as they expand the
option for optimizing magnetoresistive properties for applications
like spintronic devices with low dissipation. The discovery of the large
enhancement of magnetoresistance in Sc-doped pyrochlores $\rm Tl_2Mn_2O_7$
\cite{shimakawa96,subramanian96,ramirez97} with the same
oxidation state of the Mn ion and essentially unchanged transition
temperature $\TC$ suggests a new paradigm for manipulating magnetoresistance.
A theoretical model attributed the CMR phenomena in
$\rm Tl_{2-x}Sc_xMn_2O_7$ to magnetic polarons formed above the $T_c$
as the carrier concentration is sufficiently
low \cite{majumdar98}. It is proposed that the magnetic fluctuations
generate an effective static potential that scatters the carriers and
gives rise to electric resistance. The fact that low-field magnetoresistance scales
well with carrier densities over two orders of magnitude and in materials
with different magnetic models reveals a ubiquitous relationship
between the magnetoresistivity and charge carrier density in
ferromagnetic metals and doped semiconductors \cite{majumdar98a}.

Despite these differences between the perovskite and pyrochlore manganites, an
essential element these materials and all other CMR materials commonly share
is magnetic polarization, which minimizes spin scattering and thus electrical
resistance. However, the ferrimagnet \mst~is a surprising exception to
this rule. The recently discovered CMR in this material is realized only when
the magnetic polarization is avoided \cite{ni21}. The inherent
frustration due to competing exchange interactions between neighboring Mn ions
prevents the formation of a long-range order until the temperature is
lowered to $\TC = 78$~K \cite{rimet81,may17,sala22}. The large
linear term in the heat capacity confirms critical magnetic fluctuations
arising from the competing exchange interactions, which may hold the key to
understanding the CMR \cite{ni21}. \mst~has a magnetic easy axis within the
basal plane and a magnetic hard axis along the $c$ axis \cite{may17}. The CMR
occurs only when the magnetic field, $H$, is applied along the $c$ axis and it
is absent when $H$ is applied within the basal plane. This specific character
was later confirmed in a separate study, in which the CMR is attributed to the
formation of the nodal-line structure of the valence Te band \cite{seo21}.
Nevertheless, an adequate understanding cannot be established
without a thorough microscopic identification of the spin structure, which
is conspicuously lacking.  In this Letter, we present a comprehensive neutron
diffraction characterization of the spin structure as functions of temperature
and magnetic field applied along the $c$ axis where the CMR takes place. This
Letter uncovers a noncollinear magnetic structure with the magnetic moments
predominately lying within the basal plane but tilting toward the $c$ axis by
$\sim 10^\circ$ and a strong spin fluctuation persistent well above $\TC$
which the electrical resistivity closely tracks. Application of magnetic field
along the $c$ axis only slowly tilts the spins toward the $c$ axis but
retains the underlying antiferromagnetic (AFM) configuration. 

Single crystals of \mst~were grown using a flux method similar to that in
Refs.~\cite{may17,liu18,ni21}.  The structural information of the crystal was
investigated using a Rigaku XtaLAB PRO diffractometer equipped with a
HyPix-6000HE detector. A molybdenum anode was used to generate x rays with
wavelength $\lambda = 0.7107~\AA$. The samples were cooled by cold nitrogen
flow provided by an Oxford N-Helix cryosystem. A single crystal with
dimensions of $2\times2\times0.5$ mm$^3$ was chosen for the diffraction study using the
HB1 triple axis spectrometer at the High Flux Isotope Reactor (HFIR) and the
CORELLI diffractometer at the Spallation Neutron Source, all at Oak Ridge
National Laboratory (ORNL). The sample was aligned in the $[H, H, L]$ and
$[H,0,L]$ scattering planes for the zero-field study using a closed cycle
refrigerator, and subsequently in the $[H, K, 0]$ scattering plane to study
the field-dependent evolution of the spin configuration using a 5-T
superconducting magnet. The same single crystal was also used for 
{\it in situ} neutron diffraction studies with in-plane current applied.

\mst~has a layered structure with trigonal symmetry (SG 163, $P\bar31c$)
\cite{vincent86,may17,liu21}. There are two inequivalent Mn$_1$ and Mn$_2$
sites in the unit cell. The $\rm MnTe_6$ octahedra are composed of Mn$_1$ ions
that form an edge-sharing honeycomb lattice in the $ab$ plane, while the $\rm
MnTe_6$ octahedra consisting of Mn$_2$ sites form a triangular-lattice
sandwiched between the honeycomb layers [Fig.~1(a)]. This leads to a
magnetically frustrated network of Mn ions with three dominant
antiferromagnetic exchange interactions \cite{may17,sala22}. The nearest neighbor (NN)
exchange interaction $J_1$ is between the Mn$_1$ ion of the honeycomb layer
and the Mn$_2$ ion of the triangular layer with bond distance 3.541(1)~\AA,
the next NN coupling $J_2$ is between the Mn ions within the basal
plane with bond distance 4.056(1)~\AA, and the third NN exchange interaction
($J_3$) is for Mn ions across the plane with bond distance 5.401(1)~\AA [all
bond lengths are measured at 250~K].  \mst~orders ferrimagnetically below $\TC
\approx 78$~K \cite{rimet81,may17}.  There is no structural transition or
obvious lattice constant anomaly across the $T_c$. Details are given in
Supplementary Material \cite{supporting}.

Although the crystal maintains the same structural symmetry across the
transition, the magnetic configuration cannot be described using any of the
four maximal magnetic space groups with the hexagonal lattice. A lower
monoclinic magnetic space group $C2'/c'$ (no.~15.89, Belov-Neronova-Smirnova
setting \cite{belov57}) correctly describes the spin configuration at a base
temperature of 5~K. The transformation from the hexagonal to monoclinic
structure is obtained by
${\bf a}_m = -{\bf a}_h,
{\bf b}_m = {\bf a}_h + 2{\bf b}_h$,
and ${\bf c}_m = -{\bf c}_h$,
where $m$ and $h$ denote the monoclinic and the hexagonal cells. Because the
twofold rotation $2'$ acting on the Mn$_2$ site results in a magnetic ion at the
same position, this symmetry operation requires a sign change in the ${\bf
b}_m$-axis spin component, and thus dictates that the in-plane component be strictly
along the ${\bf a}_m$ or $[1,0,0]$ in the hexagonal setting. There is no constraint for the
${\bf b}_m$ component for Mn$_1$. In addition, nonzero ${\bf c}_m$ components
are allowed at both sites. The magnetic structure of \mst~was previously
reported to be a collinear spin order; the moments reside in the basal plane
forming a ferrimagnetic order with strong easy-plane anisotropy \cite{may17}.
However, the magnetization measurement with field $H\parallel c$ shows a
continuous growth of the intensity suggesting a rather soft spin component in
the $c$ axis, although the anisotropy field is approximately 13~T
\cite{ni21}. A strong angular-dependent CMR with seven orders of magnitude drops
in resistivity in \mst, which was first reported in
Ref.~[\onlinecite{ni21}] and later confirmed in Ref.~[\onlinecite{seo21}],
prompt a thorough reexamination of the spin configuration.

\begin{figure}[htp]
 \includegraphics[width=3.2in]{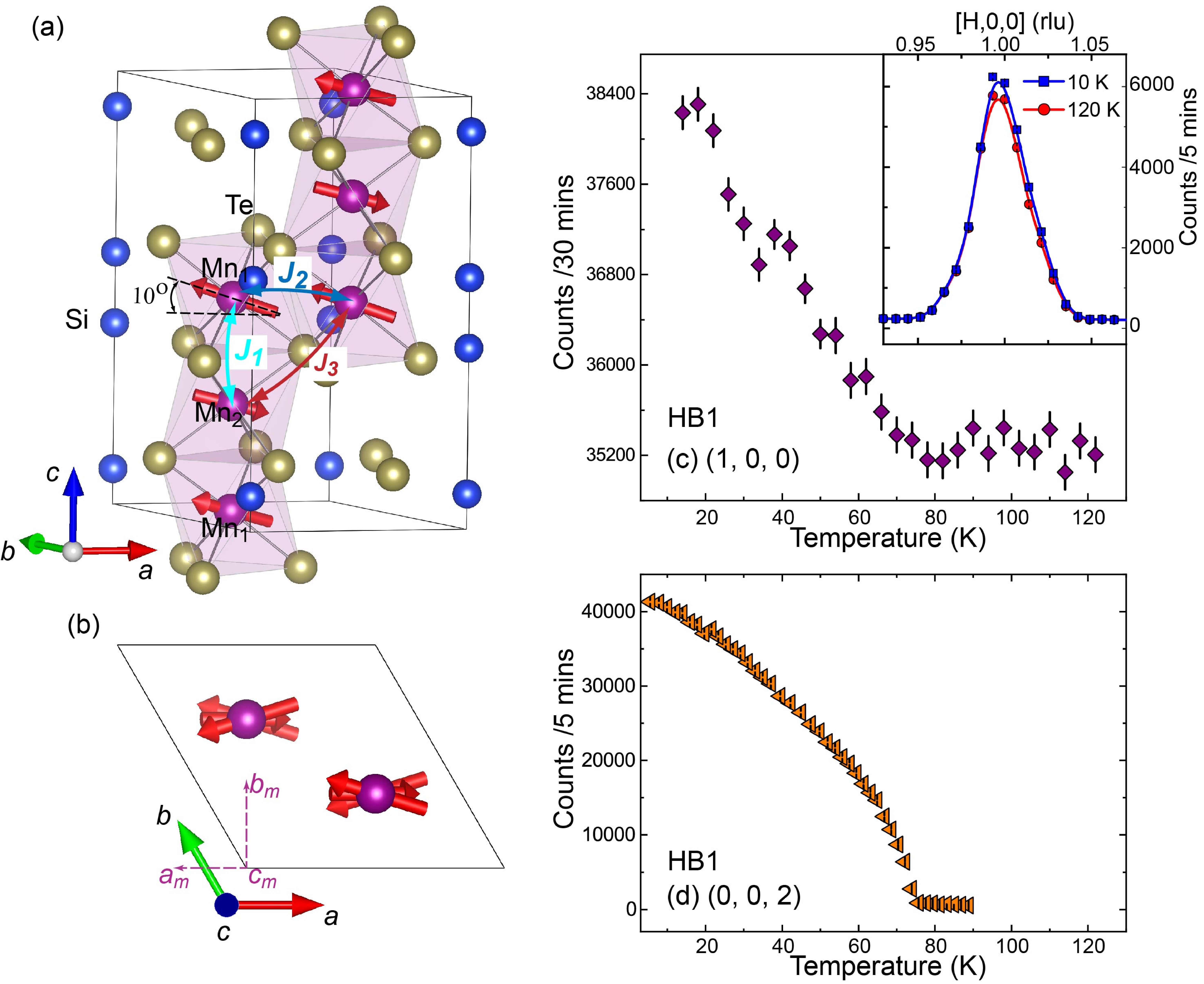}
 \caption{(a) The refined zero-field magnetic structure of \mst~at 5 K. Two
    symmetry inequivalent Mn sites are located at $(1/3,2/3,z)$ with $ z
    \approx 0$ and $1/4$. The spin directions are dominantly along the
    $[1,0,0]$ direction and $\sim 10^\circ$ tilting away from the basal plane.
    The pathways of the first three AFM exchange
    interactions $J_1 \sim -402$, $J_2 \sim -73$, and $J_3 \sim -172$~K/Mn
    \cite{may17} between neighboring magnetic ions are also shown in cyan,
    blue, and red arrows.  (b) The magnetic structure projected in the basal
    plane. $a_m$, $b_m$, and $c_m$ are the axes in the monoclinic setting. (c)
    The $T$ dependence of the $(1,0,0)$ peak associated with the canted spin
    configuration. Inset shows the comparison of the wave-vector scan at 10 
    and 120 K across the $(1,0,0)$ Bragg peak. (d) The $T$ dependence of the
    dominant $(0,0,2)$ magnetic peak for the ferrimagnetic spin order.
 }
\end{figure}

Figure 1(c) shows the temperature ($T$) dependence of the $(1,0,0)$ reflection
that clearly exhibits an enhancement in amplitude below $\TC \sim 78$~K. A
detailed comparison of wave-vector scans through the same peak at 10 and 120~K
further confirms the extra magnetic scattering intensity is about 10\% of the
nuclear one.  The coherent magnetic scattering intensity probes the spin
component ${\bf S}_{\perp}$ perpendicular to the momentum transfer ${\bf Q}$,
${\bf S}_{\perp}=\hat{\bf Q}\times ({\bf S} \times \hat{\bf Q})$.  The fact
that additional intensity at ${\bf Q}=(1,0,0)$ appears below the transition
indicates the presence of a spin component perpendicular to the $[1,0,0]$
vector. This could suggest a canted spin configuration which can be
quantitatively characterized from the refinement. To do so, we perform a full
360$^\circ$ map at 5, 100, and 200~K using the white beam single crystal
diffractometer CORELLI \cite{ye18}.  The crystal was rotated along the
vertical axis with 2$^\circ$ steps between runs. Each reflection is collected
with averaged redundancy of 5.  Proper Lorentz, time-of-flight spectrum,
and detector efficiency corrections are applied by the method detailed in
Ref.~[\onlinecite{michels-clark16}]. A total of $\sim 500$ reflections
are obtained for a simultaneous refinement for the magnetic and nuclear
structure using the JANA2006 program \cite{petricek14}. Table I shows the refined
parameters; the final spin structure is illustrated in Fig.~1(a). Both Mn
sites form  $\sim 10^\circ$ canting angles toward the $c$ axis. The nearly
antiparallel Mn1 and Mn2 spins imply a considerable AFM interaction in
between.  Since the in-plane component of Mn$_1$ does not have the same
constraint as Mn$_2$, the projected view in the $ab$ plane shows a
noncollinearity of the overall spin configuration [Fig.~1(b)].

\begin{table}[ht!]
  \caption{Refined magnetic moments of symmetry independent atoms. Mn$_1$ is
  located at $(1/3, 2/3, z)$, and Mn$_2$ is located at $(1/3,2/3,1/4)$ in SG
    163, $P\bar31c$.  The magnetic space group $C2'/c'$ puts no constraint on
    Mn$_1$ but $m_b=0$ for Mn$_2$. $m_a$, $m_b$, $m_c$, and $|m|$ denote
    the amplitude of spin components along the monoclinic $a_m$, $b_m$, and
    $c_m$ axes, and the total moment, respectively.  Three twin domains are
    present with a volume ratio of 36:34:30.
    }
  \label{tab:BV}
\begin{ruledtabular}
\begin{tabular}{cccccc}
Label & Multiplicity & m$_a$ & m$_b$ & m$_c$ & $|m|$ \\ \hline
Mn$_1$ & 4 & -3.42(9) & 1.50(8) & 0.95(3) & 4.55(3)\\
Mn$_2$ & 2 & 4.13(8) & 0 & -0.73(3) & 4.20(3) \\
 \end{tabular}
\end{ruledtabular}
\end{table}

More details on the magnetic structure are characterized near and above the
transition.  In contrast to the long-range order showing sharp magnetic
reflections in the $(H,H,L)$ plane [Fig.~2(a)], notable magnetic diffuse
scattering exists at the magnetic peaks $(0,0,2)$, and $(1,1,2)$ and equivalent
reflections where the Miller indices are dominant by $L$ component
[Fig.~2(b)]. This reveals a considerable in-plane spin fluctuation that can
only be probed by the momentum transfer perpendicular to the spin component.
The in-plane magnetic fluctuations are slightly broadened compared to those
along the $c$ axis yet maintain a three-dimensional character.
Surprisingly, the diffuse scattering is present over a broad range of
temperature near the transition. It peaks at the transition and persists
well above $\TC$ (Supplementary Material \cite{supporting}). 
As shown in Fig.~2(d), peak intensities collected at ${\bf
Q}=(-0.025, 0, 2)$ exhibit a slow decay as the temperature is raised.  The
intensity remains finite at temperatures close to $\sim 2\TC$. The
wave-vector-dependent susceptibility above the transition can be described in
the Ornstein-Zernike form $\chi(q) =1/(q^2+k^2)$,where $k=1/\xi$ is the
inverse correlation length. The in-plane correlation length $\xi_{ab}(T)$
displays a similar suppression on warming from 14.5(5)~\r{A} at 85 K to 4(1)
\r{A} at 180 K. These values are rather small and comparable with the in-plane
lattice parameter, highlighting the confinement of the magnetic clusters above
the transition. Interestingly, the thermal evolution of the $\xi^2$ overlaps
nicely with that of the electric resistivity, which has important implications
for the magnetoresistance in that temperature regime as discussed below.

\begin{figure}[htp]
  \includegraphics[width=3.4in]{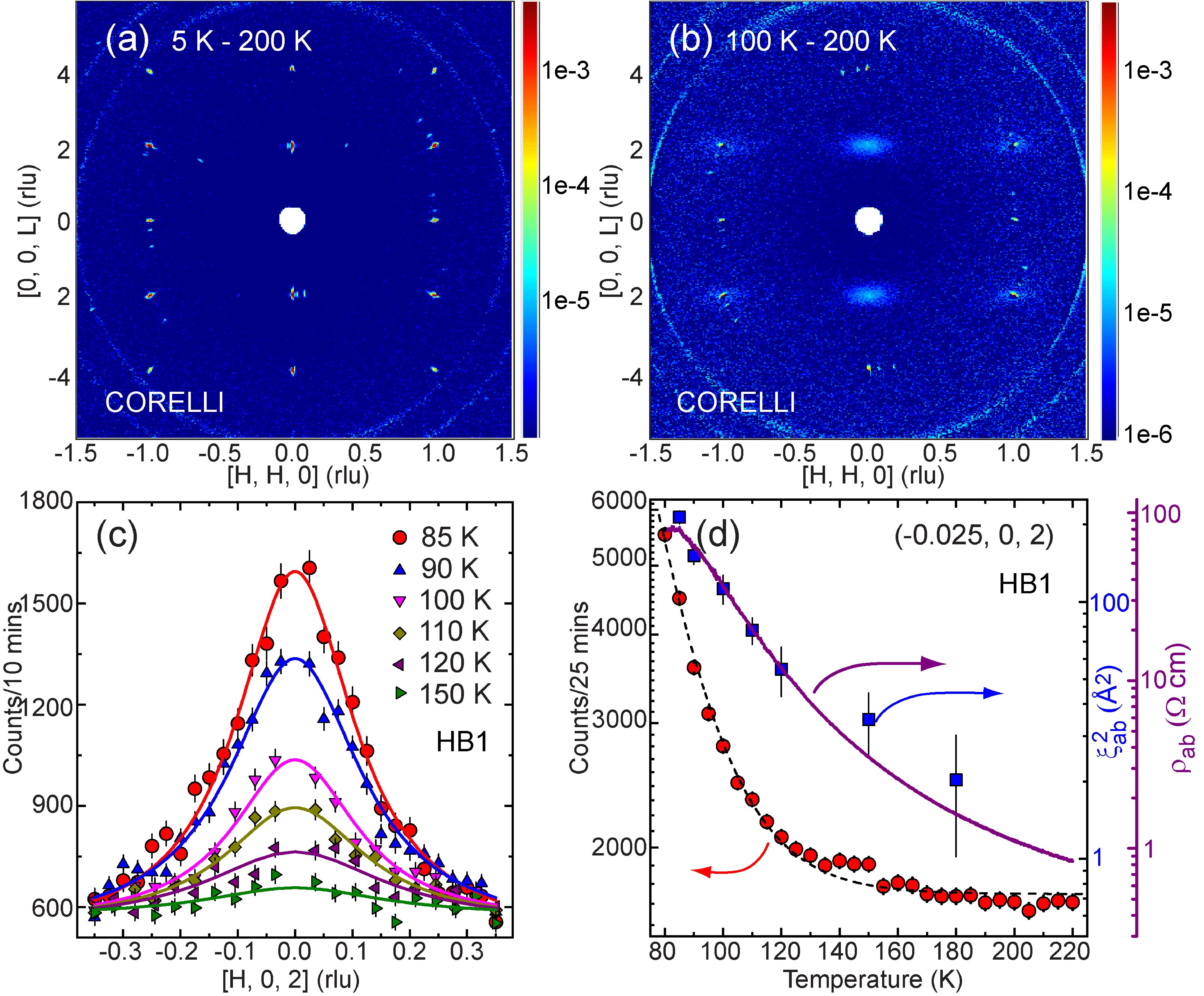}
  \caption{ (a) The contour plot of the neutron diffraction data in the
  $(H,H,L)$ scattering plane collected on CORELLI at 5 K with 200 K data subtracted
  as background. (b) The similar plot at 100 K.
  (c) Wave-vector scans along the [1,0,0] direction across the magnetic (0,0,2) peak
  at selected temperatures. The solid lines are the Lorentzian fits to the data points.
  (d) (left) The $T$ dependence of the peak intensities at ${\bf Q}=(-0.025,0,2)$.
  The dashed line is the fit to the data points using exponential decay form
  $I(T)=A_1e^{-(T-T_0)/t_1}+I_0$, with $T_0=79(1),t_1=16.3(5), A_1=3800(20)$,
  and $I_0=1720(15)$. (right) the thermal evolution of in-plane
    resistivity $\rho_{ab}(T)$ (solid purple line) and $\xi_{ab}^2(T)$, where
    $\xi_{ab}$ (blue solid square) is in-plane correlation length.
  }
\end{figure}

We now turn to the response of the spin structure with magnetic field applied along the
$c$ axis. The full data sets are collected at 0, 1, and 4.75~T to
determine the spin order, while more detailed field dependence studies
are performed focusing on representative reflections of $(0,1,1)$ and
$(0,1,0)$. As shown in Figs.~3(a)-3(b), both peak intensities exhibit abrupt
enhancement in the small field regime ($<1$~T). The refinement at
1~T indicates a melting of the domain walls between three
nearly equally populated magnetic twins, leading to a single domain with no
change in the magnetic structure.  More importantly, the data indicate that
the Mn$_1$ ion carries a magnetic moment of 4.5 $\mu_B$ whereas the moment at
the Mn$_2$ site is 4.2 $\mu_B$. Therefore, the net magnetic
moment, according to the magnetic configuration in Fig.~3(a), is 1.6
$\mu_B$/Mn (equal to $[2\times4.5 - 4.2]/3$).
This is perfectly consistent with the saturated magnetization of 1.56
$\mu_B$/Mn in the basal plane at $\mu_0 H > 0.1$~T \cite{ni21,seo21}.
We note there is a slight misalignment of the crystal orientation where
the applied field deviates from the crystalline $c$ axis by 3$^\circ$, which
translates into an in-plane field $\sin(3^\circ) \times 1$~T=0.05~T (close to
0.1~T) that overcomes the coercive field.

Further increasing the magnetic field along the $c$ axis systematically suppresses the
intensity of both $(0,1,1)$ and $(0,1,0)$ reflections [Figs.~3(a)-3(b)].  Such
behavior is distinct from that of the low-field regime and signals a gradual
change in the spin configuration once the system enters the single domain
state. In particular, the refinement reveals a further tilting of all spins
toward the $c$ axis from 10$^\circ$ at 1~T to 33$^\circ$ at 4.75~T
(supplementary material \cite{supporting}). 
This is illustrated by the solid lines in Figs.~3(a)-3(b) showing evolution
of the calculated intensities with spin moment tilting away from the basal
plane, which capture the trend reasonably well.  The in-plane lattice
parameter $a$ shows a monotonic increase with field applied along the $c$
axis.  An accurate determination of the field dependence of the $c$ lattice
parameter is not possible since the crystal is oriented with the $c$ axis
aligned out of the plane. We also measured the thermal evolution of both
magnetic reflections at $\mu_0 H = 5$~T. In contrast to the sharp anomaly
observed in zero field, the transition is significantly broadened. This is
expected for a ferrimagnet, because the applied magnetic field forces the
spins to be aligned with $H$ above the intrinsic magnetic transition.

\begin{figure}[!ht]
  \includegraphics[width=3.2in]{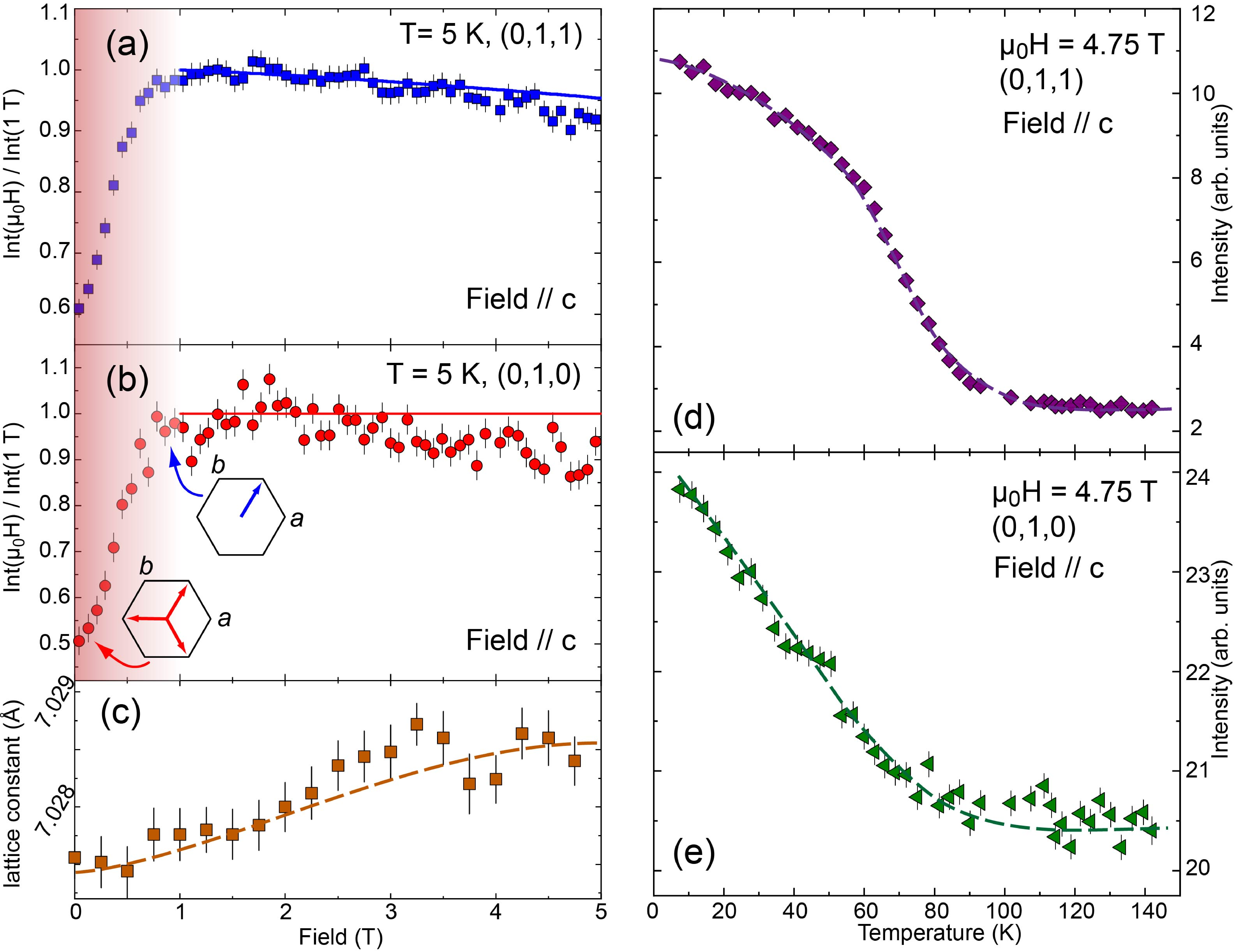}
  \caption{ Field dependence of the (a) $(0,1,1)$ and (b) $(0,1,0)$ magnetic
    reflections with field direction parallel to the $c$ axis.
    Solid blue and red lines for $\mu_0 H > 1$~T are the
    calculated intensities with angle tilting smoothly from 10$^\circ$ to
    $33^\circ$. Inset of (b) illustrates the conversion of
    three equally populated magnetic twins into a single domain at field
    $\sim$ 1~T. (c) The field dependence of the in-plane lattice constant
    $a$ at 5 K.  The temperature dependence of peak intensity of the (d)
    $(0,1,1)$ and (e) $(0,1,0)$ peaks at field of 4.75~T.  Dashed lines
    are guides to the eye.
  }
\end{figure}

The magnetic response to the applied magnetic field described above make the
observed CMR with seven orders of magnitude drops in resistivity in
\mst~indeed exceptional \cite{ni21}. Seo {\it et al.}~propose that the main
mechanism driving an insulator-metal transition is the magnetic valve effect,
where the spin rotation by external fields drastically reduce the electronic
band gap and the charge conduction \cite{seo21}. In light of the neutron
diffraction data presented in Figs.~1-3, such a proposal becomes inadequate to
explain the observed CMR.  This is because our neutron diffraction results
reveal that \mst~already possesses a canted angle of 10$^\circ$ away from the $ab$
plane at ambient condition, and the canting angle changes merely to $33^\circ$
at the applied field of 4.75~T along the $c$ axis.  Yet, the CMR abruptly
takes place at a critical field $\mu_0 H_c=3$~T, much smaller than the
saturation field of 13~T transforming spins to a fully polarized state
[Fig.~4(a)]. The lack of a parallel response to $H$ in $\rho_{ab}$ and the
magnetic canting confirms that the magnetic spins alone cannot account for the
observed CMR.  Furthermore, a dramatic change in the electric conductivity is
observed with application of an in-plane electric current [Fig.~4(b)]. With
increasing current from 0.1 to 4~mA, the resistivity at 6~K drops by half, yet
the {\it in situ} neutron diffraction measurement shows disproportionally small
change in magnetic order. The intensity of the magnetic (0,0,2) peak reduces
by 15\% while $\rm T_c$ shifts only slightly to 75~K at 4~mA.  More dramatic
change in the transport properties and complex phase transitions are observed
under small in-plane direct current density \cite{electricnote}.
These observations clearly imply that magnetoresistance in \mst~is rather
complicated and different mechanisms have to be invoked.

\begin{figure}[!ht]
  \includegraphics[width=3.4in]{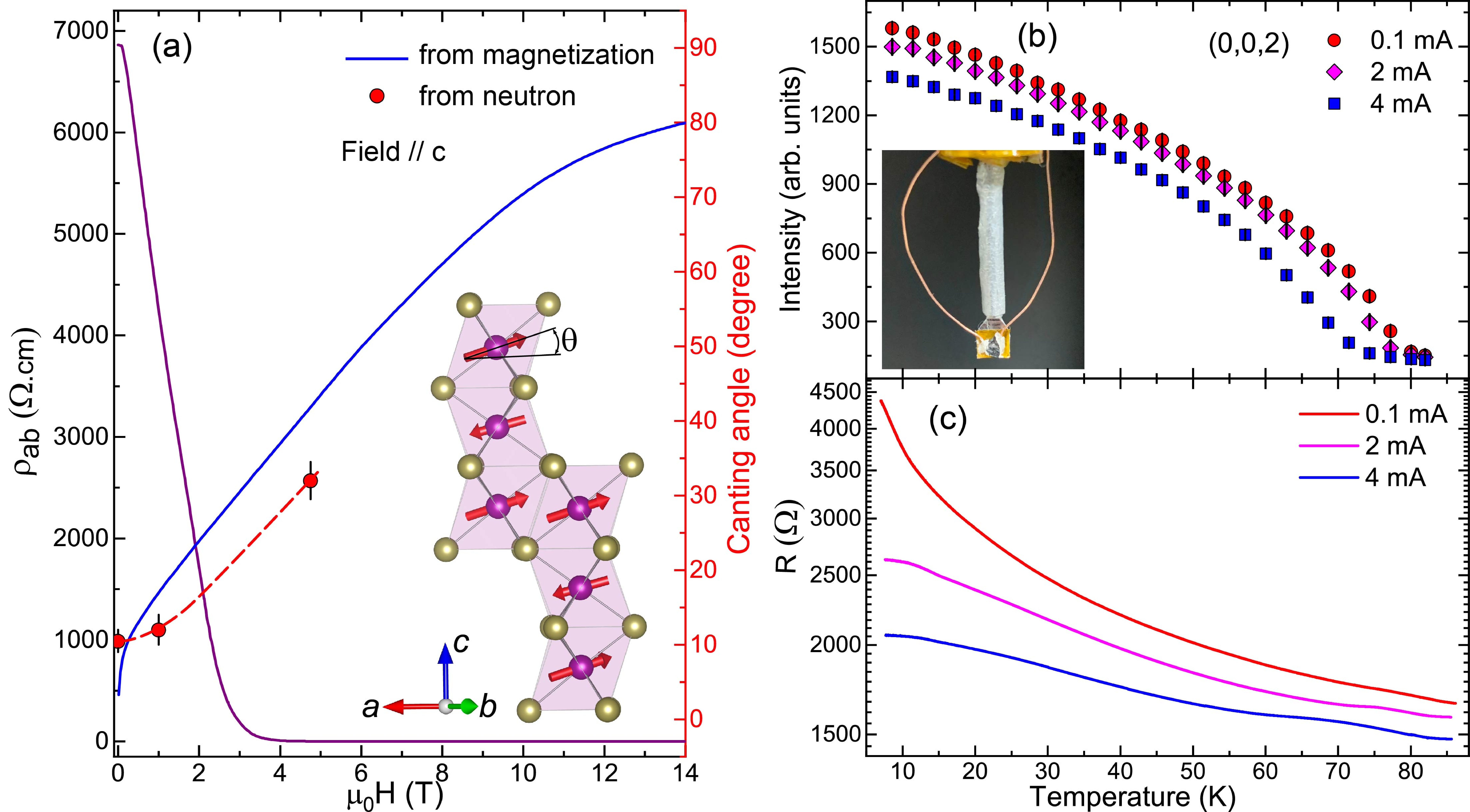}
    \caption{
    (a) Field dependence of the in-plane resistivity $\rho_{ab}$ (purple) and
    canted angle obtained from neutron diffraction and magnetization measurements.
    (b) $T$ dependence of the $(0,0,2)$ magnetic reflection as
    function of increasing in-plane electric current. (c) The {\it
    in situ} measurement of resistance under the same experimental
    configuration. Inset shows the setup that performs simultaneous
    neutron diffraction and electric transport measurement. }
\end{figure}

In the low carrier density semiconductors $\rm Tl_{2-x}Sc_xMn_2O_7$
\cite{shimakawa96,subramanian96,ramirez97} and $\rm Eu_{1-x}Gd_xSe$
\cite{vonmolnar67}, the coupling between itinerant electrons and the
fluctuating magnetic spins are emphasized as the main source contributing to
resistance in the regime above $\TC$ \cite{majumdar98,majumdar98a}.  According
to the model, the magnetotransport in the low carrier concentration limit ($n\leq
1/\xi^3$, with $\xi$ the magnetic correlation length) ferromagnet is governed
by the scattering process of the electron gas coupled to spin fluctuation. The
resistivity $\rho$, can be derived as $\rho \sim \xi^2/(1+\xi^2q^2)$ in the
Ornstein-Zernike approximation, with $q$ being the momentum transfer of the
ferromagnetic fluctuations \cite{fisher68}. It is evident that $\rho \sim \xi^2$
for the low electron density situation over a wide range of temperature above
the transition. The strength of the coupling constant $C$ between the
itinerant electrons and local moment can be estimated
from $\Delta \rho(H)/\rho = C(m(H)/m_{\rm sat})^2$, where $m(H)$ and $m_{\rm
sat}$ are field-induced and saturation magnetization, and usually exhibits
anomalous large values in the fluctuation scenario.

Indeed, \mst~displays a number of characteristics of the magnetic
semiconductors exhibiting CMR driven by spin fluctuation; the charge carrier
density is low at 10$^{-4}$ per formula unit,
which is similar to $\rm Tl_2Mn_2O_7$ ranging from 0.001 to0.005 \cite{ramirez97}, 
but significantly smaller than in doped $\rm
La_{1-x}Sr_xMnO_3$ \cite{ramirez97a}. The constant $C$ is 105 above
$\TC$ \cite{ni21}, nearly two orders of magnitude higher than those in the
metallic manganites \cite{tokura94}. Third, just like the magnetoresistivity
that extends to temperature much higher than the transition, the magnetic
diffuse scattering also persists at high temperature [Fig.~2(d)] revealing
a close connection between the two. The square of magnetic correlation
lengths of the fluctuating spins follows the thermal evolution of the
resistivity. This is in line with the theoretical model that $\rho\sim \xi^2$
even with $d \rho/d T<0$ \cite{majumdar98,majumdar98a} and the
observation of a polaronic transport behavior above the transition
\cite{liu21}. Finally, the density function theory calculation
\cite{may17} reveals that the magnetic exchange interactions are dominated by
the strongest nearest neighbor coupling $\rm J_1$ along the $c$ axis, followed
by two weaker terms $\rm J_2$ and $\rm J_3$ perpendicular to the $c$ axis
[inset of Fig.~1(a)]. Such calculation was recently verified by
spin wave excitation study using inelastic neutron scattering,
which further reveals exchange anisotropy due to the spin-orbit coupling at
the Mn1 site \cite{sala22}. The system is frustrated because the first
three magnetic exchange constants are of the same AFM type.  This explains
the rather low transition $\TC \sim 78$~K comparing to a much higher
Curie-Weiss temperature $\Theta_{CW}\sim -277$~K and the prevailing magnetic
fluctuation.  With magnetic field applied along the $c$ axis, the in-plane
lattice constant shows gradual expansion upon increased field [Fig.~3(c)].
This in-plane lattice expansion could result from the magnetostriction 
effect and shows an opposite trend comparing to temperature dependence of the
lattice constant in zero applied field \cite{may17}, which conceivably  might
lead to modification of existing magnetic exchange interactions and eventual
suppression of the magnetic frustration.

In summary, our neutron diffraction study reveals a canted spin configuration
at 5 K that is crucial to understanding the unconventional CMR in the
ferrimagnetic \mst. A slow and gradual spin rotation toward the $c$ axis
occurs with increasing magnetic field applied along the same direction. Our
results provide direct evidence that the modification in spin order is not
sufficient to explain the giant magnetoresistance response when the field is
applied along the spin hard axis, and prompt further experimental
\cite{wang22} and theoretical efforts to describe the novel magnetotransport
behavior. On the other hand, the prominent short-range magnetic diffuse
scattering near $T_c$, the close connection between the square of spin-spin
correlation lengths and electric resistivity, strongly suggests the relevance
of fluctuating moments near the magnetic transition contributing to the
transport properties in this low carrier density semiconductor. A future
single crystal magnetic diffuse scattering study under applied
field is highly desirable to provide important insight into the role of the
fluctuating spins. 

We thank P.~Majumdar and P.~Littlewood for stimulating discussion. Research
at ORNL's HFIR and SNS was sponsored by the Scientific User Facilities
Division, Office of Basic Energy Sciences, U.S.~Department of Energy.
Work at the University of Colorado was supported by NSF via Grants No.~DMR
1903888 and DMR 2204811.

\vfill
\bibliographystyle{h-physrev}

\end{document}